\documentclass[11pt]{article}

\usepackage[margin=1.25in]{geometry}
\usepackage{url}            
\usepackage{authblk}
\usepackage{amsfonts}       
\usepackage{graphicx}
\usepackage{amsmath}
\usepackage{amsthm}
\usepackage[ruled]{algorithm2e}
\usepackage{algorithmic}
\usepackage{caption}
\usepackage{soul}
\usepackage{subcaption}
\usepackage[colorlinks=true,citecolor=blue, linkcolor=blue]{hyperref}

\usepackage[numbers,compress]{natbib}
\usepackage{array}
\usepackage{color}

\newtheorem*{theorem*}{Theorem}

\theoremstyle{definition}

\renewcommand{\a}{{\bf a}}
\renewcommand{\b}{{\bf b}}
\newcommand{\bfc}{{\bf c}}

\newcommand{\h}{{\bf h}}

\newcommand{\s}{{\bf s}}
\renewcommand{\v}{{\bf v}}
\newcommand{\w}{{\bf w}}
\newcommand{\x}{{\bf x}}
\newcommand{\y}{{\bf y}}
\newcommand{\z}{{\bf z}}

\newcommand{\A}{{\bf A}}
\newcommand{\B}{{\bf B}}
\newcommand{\C}{{\bf C}}

\newcommand{\I}{{\bf I}}
\newcommand{\M}{{\bf M}}

\newcommand{\R}{\mathbb{R}}

\newcommand{\W}{{\bf W}}

\newcommand{\bxi}{\boldsymbol{\xi}}
\newcommand{\bzeta}{\boldsymbol{\zeta}}

\DeclareMathOperator{\tr}{Tr}

\title{Normative framework for deriving neural networks with multi-compartmental neurons and non-Hebbian plasticity}

\author[1,5]{David Lipshutz\thanks{Equal contribution.}$^{1}$, Yanis Bahroun$^\ast$$^{1,2}$, Siavash Golkar$^\ast$$^{1}$, \vspace{3pt}\\Anirvan M. Sengupta$^{2,3,4}$, and Dmitri B.\ Chklovskii}
\affil[1]{Center for Computational Neuroscience, Flatiron Institute}
\affil[2]{Center for Computational Mathematics, Flatiron Institute}
\affil[3]{Center for Computational Quantum Physics, Flatiron Institute}
\affil[4]{Department of Physics and Astronomy, Rutgers University}
\affil[5]{Neuroscience Institute, NYU Langone Medical Center}

\begin{document}

\maketitle

\begin{abstract}
An established normative approach for understanding the algorithmic basis of neural computation is to derive online algorithms from principled computational objectives and evaluate their compatibility with anatomical and physiological observations. Similarity matching objectives have served as successful starting points for deriving online algorithms that map onto neural networks (NNs) with point neurons and Hebbian/anti-Hebbian plasticity. These NN models account for many anatomical and physiological observations; however, the objectives have limited computational power and the derived NNs do not explain multi-compartmental neuronal structures and non-Hebbian forms of plasticity that are prevalent throughout the brain. In this article, we unify and generalize recent extensions of the similarity matching approach to address more complex objectives, including a large class of unsupervised and self-supervised learning tasks that can be formulated as symmetric generalized eigenvalue problems or nonnegative matrix factorization problems. Interestingly, the online algorithms derived from these objectives naturally map onto NNs with multi-compartmental neurons and local, non-Hebbian learning rules. Therefore, this unified extension of the similarity matching approach provides a normative framework that facilitates understanding multi-compartmental neuronal structures and non-Hebbian plasticity found throughout the brain.
\end{abstract}

\section{Introduction}

Advances in theoretical neuroscience are often driven by the development of normative frameworks that explain physiological and anatomical observations from the perspective of computational principles \citep{attneave1954some,barlow1961possible,srinivasan1982predictive,oja1982simplified,atick1992does,van1992theory,olshausen1997sparse,rao1999predictive,chen2006wiring,pehlevan2015hebbian,mlynarski2021efficient}.
These top-down frameworks start with computational objectives from which physiological and anatomical implications are derived and compared with experimental observations. 
In the context of understanding the algorithmic basis of neural computation, this approach involves starting with a computational objective, deriving an online algorithm that can be implemented in a neural network (NN), and comparing the NN model with experimental observations.

\begin{figure}
    \centering
    \includegraphics[width=.6\textwidth]{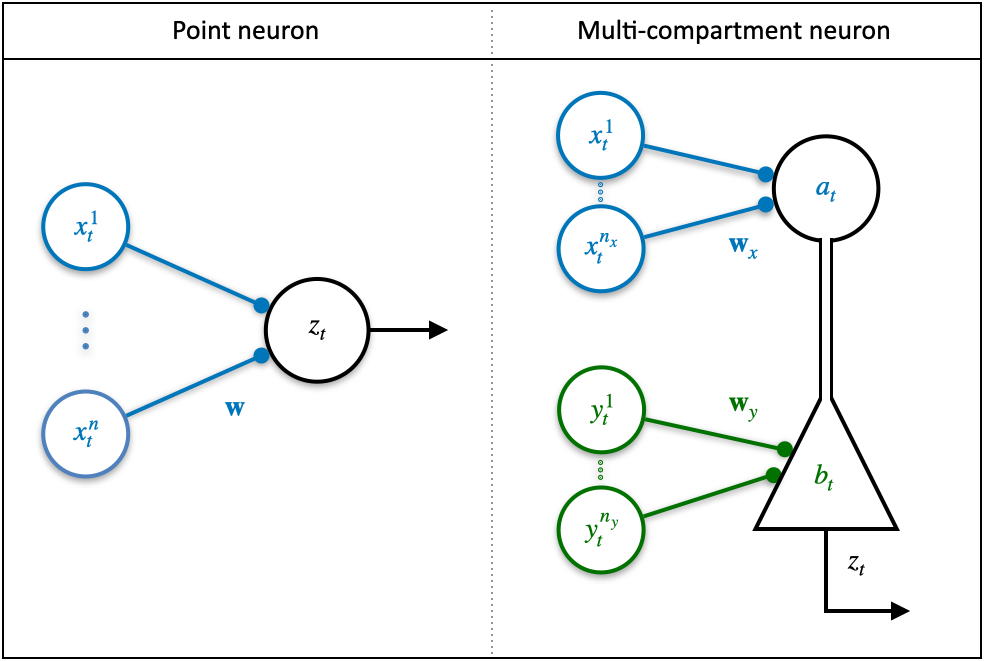}
    \caption{Left: Oja's point neuron with inputs $\x_t$, synaptic weights $\w$, output $z_t=\w\x_t$ and Hebbian plasticity. Right: Multi-compartment model of a pyramidal neuron with separate distal compartment (circular region) and proximal compartment (triangular region). Feedback inputs $\x_t$ target the distal portion of the apical tuft to generate the current $a_t=\w_x\x_t$, which drives non-Hebbian plasticity at the proximal synapses $\w_y$. Feedforward inputs $\y_t$ target the proximal region to generate the current $b_t=\w_y\y_t$. The output $z_t$ is a function of the currents $a_t$ and $b_t$ and, depending on the model, is sometimes represented in a third compartment.}
    \label{fig:pca}
\end{figure}

In a pioneering example of this approach, \citet{oja1982simplified} proposed an online algorithm for Principal Component Analysis (PCA) \citep{pearson1901liii}, a popular unsupervised dimensionality reduction method, which can be implemented in a point neuron (i.e., a neuron that only represents its scalar output) with Hebbian plasticity, Figure \ref{fig:pca} (left). Hebbian plasticity, named after Hebb \cite{hebb1949organisation}, refers to synaptic updates that are proportional to the product of the pre- and postsynaptic neural outputs. Experimental evidence of Hebbian plasticity came with the discovery of long term potentiation \cite{bliss1973longA,bliss1973longB} and since then a variety of forms of Hebbian plasticity have been observed \citep{caporale2008spike}. Oja's model of a point neuron thus offers a link between experimentally observed Hebbian plasticity and an unsupervised learning objective. However, in the few decades following Oja's work, efforts to extend Oja's approach to extract multiple principal components resulted in NNs that used non-local learning rules \citep{oja1983analysis,sanger1989optimal,leen1990hebbian}.

Building on Oja's seminal work, Pehlevan, Chklovskii et al.\ recently developed a normative framework to extract multiple principal components using similarity matching objectives \citep{pehlevan2015hebbian,pehlevan2019neuroscience,deepSM,tecsileanu2022neural,genkin2022transformation}, which minimize the difference between the similarity of the NN inputs and the similarity of the NN outputs. Starting from these objectives, they derived online algorithms that map onto multi-channel NNs with point neurons and Hebbian plasticity. This normative framework proved useful for linking unsupervised learning objectives to Hebbian plasticity and several anatomical and physiological observations \citep{pehlevan2017clustering,bahroun2019similarity,benna2021place,chapochnikov,drift,lipshutz2022interneurons}. However, the similarity matching objectives have limited computational power and the derived NNs cannot explain multi-compartmental neuronal structures and other forms of synaptic plasticity prevalent throughout the brain \citep{magee2020synaptic}.

Most neurons in the brain have multi-compartmental structures and employ intricate forms of non-Hebbian plasticity. In particular, these neurons represent biophysical quantities beyond their output, such as local dendritic currents, and these quantities constitute key variables in the neurons' synaptic learning rules. For example, pyramidal neurons---the primary excitatory neurons of the cortex capable of performing complex computations \citep{gidon2020dendritic}---receive inputs to their proximal and distal dendrites from distinct neural populations and integrate these inputs in separate compartments \citep{spruston2008pyramidal}, Figure \ref{fig:pca} (right). Integrated distal inputs generate calcium plateau potentials that drive non-Hebbian plasticity in the proximal dendrites \citep{takahashi2009pathway}. What are the computational objectives that lead to these more complex neuronal structures and intricate forms of non-Hebbian plasticity?

In a series of recent works \citep{lipshutz2021biologically,lipshutz2020biologically,bahroun2021normative,golkar2022cpca,lipshutz2022single}, we have extended the similarity matching framework to include objectives for more complex learning tasks. Examples include computational objectives for Canonical Correlation Analysis (CCA), Slow Feature Analysis (SFA), Independent Component Analysis (ICA) and contrastive PCA* (cPCA*), which can respectively be interpreted as linear instantiations of the following computational principles: associative learning of multi-modal inputs, learning temporally invariant features, redundancy reduction and contrastive learning. Interestingly, the algorithms derived from these objectives naturally map onto NNs with multi-compartmental neurons and local, non-Hebbian forms of plasticity. Therefore, these works offer a potential normative account of these anatomical and physiological observations.

In this article, we provide a unified framework that encompasses and generalizes these normative models of NNs with multi-compartmental neurons and non-Hebbian plasticity. 
In particular, we derive an online algorithm for solving a large class of symmetric generalized eigenvalue problems---which includes CCA, SFA, ICA and cPCA* as special cases---that establishes a precise link between synaptic plasticity rules and computational objectives.
In one direction, this framework can be used to derive NNs for solving other symmetric generalized eigenvalue problems \citep{cunningham2015linear,ghojogh2019eigenvalue}.
Conversely, given an experimentally observed non-Hebbian synaptic plasticity rule, this framework can potentially be used to predict a guiding computational objective.
Therefore, we believe this unified framework will facilitate further development of NNs for solving other relevant learning tasks and advance our understanding of NNs with multi-compartmental neurons and non-Hebbian plasticity.

The remainder of this work is organized as follows. We first review prior theoretical results on NNs with point neurons and Hebbian plasticity (section \ref{sec:Hebbian}) and experimental results on multi-compartmental neurons and non-Hebbian plasticity (section \ref{sec:non-Hebbian}). We then present a unified objective for solving a large class of symmetric generalized eigenvalue problems (section \ref{sec:gpsp}). Starting from this objective, we derive an online algorithm for solving the objective (section \ref{sec:derivation}) and show that for several examples the algorithm maps onto NNs with multi-compartmental linear neurons and non-Hebbian learning rules (section \ref{sec:nns}). 
Finally, by modifying the starting objective, we transform the problem from a symmetric generalized eigenvalue problem to a nonnegative matrix factorization problem, resulting in NNs with rectified neural outputs (section \ref{sec:nn_sm}).

\section{Hebbian NNs for unsupervised dimensionality reduction}\label{sec:Hebbian}

Early sensory processing significantly reduces the dimensionality of the inputs \citep{atick1990towards,ganguli2012compressed}. For example, the human retina is a highly convergent pathway with more than 100 fold reduction in dimensionality from photoreceptors to retinal ganglion cells \citep{hubel1995eye}. Therefore, NNs that perform unsupervised dimensionality reduction may be useful models of early sensory processing. 



\subsection{Oja's neuron for PCA}

In a seminal work, \citet{oja1982simplified} modeled a single neuron with a PCA algorithm, Figure \ref{fig:pca} (left), which can be derived as a stochastic gradient descent minimizing a reconstruction error objective \citep{yang1995projection}. At each time point $t$ the neuron receives $n$ inputs, whose activities are encoded in the column vector $\x_t$. The inputs are multiplied by the corresponding synaptic weights, which are encoded in the row vector $\w$, and summed to generate the neuron's scalar output $z_t=\w\x_t$. These synaptic weights are then updated according to the following plasticity rule, referred to as Oja's rule:
\begin{align*}
    \w\gets\w+\eta(z_t\x_t^\top-z_t^2\w),
\end{align*}
where $\eta>0$ denotes the learning rate for the synapses. After multiple iterations, the synaptic weights $\w$ converge to the principal eigenvector of the input covariance matrix $\C_X:=\langle\x_t\x_t^\top\rangle$ \citep{oja1982simplified,duflo2013random,chou2020ode}. The first term in the synaptic update, $z_t\x_t^\top$, is the product of the pre- and postsynaptic activities, so it is referred to as Hebbian plasticity. The term $-z_t^2\w$, which is proportional to the synaptic weights, can be viewed as form of homeostatic plasticity that prevents the synaptic weights from diverging.

\subsection{Hebbian NNs derived from similarity matching objectives}

Following Oja's work, several extensions to multi-channel NNs were proposed. In one line of work, online algorithms were derived from principal subspace projection (PSP) objectives and mapped onto single-layer neural networks \citep{oja1983analysis,sanger1989optimal,leen1990hebbian}; however, the synaptic updates in these NNs are not \textit{local}---they depend on variables that are not represented in the pre- or postsynaptic neurons---so they violate basic biophysical constraints. In another line of work, NNs with local, Hebbian learning rules for synapses were proposed \citep{foldiak1989adaptive,rubner1989self,rubner1990development,leen1991learning}; however, the synaptic updates were postulated rather than derived from a principled objective, so the NNs are not normative and lack theoretical understanding.

To enjoy the benefits of both of these two lines of work, a normative approach and local learning rules, \citet{pehlevan2015hebbian,pehlevan2018similarity} introduced the following similarity matching objective from multidimensional scaling~\citep{cox2008multidimensional}:
\begin{align}\label{eq:objective_psp}
    \min_{\z_1,\dots,\z_T\in\R^k}\frac{1}{T^2}\sum_{t=1}^T\sum_{t'=1}^T\left(\x_t^\top\x_{t'}-\z_t^\top\z_{t'}\right)^2,
\end{align}
where $\z_t$ denotes the output of the NN at time $t$. The objective minimizes the squared difference between the similarity of the inputs and the similarity of the outputs, where similarity is measured in terms of inner products. The optimal solution $\z_1,\dots,\z_T$ of objective \eqref{eq:objective_psp} is the projection of the inputs $\x_1,\dots,\x_T$ onto their $k$-dimensional principal subspace, i.e., the subspace spanned by the top $k$ principal components. Starting from this objective, \citet{pehlevan2015hebbian,pehlevan2018similarity} derived an online gradient-based algorithm for PSP, Algorithm \ref{alg:pca} (see section \ref{sec:derivation} for a detailed derivation).

\vspace{10pt}
\begin{algorithm}[H]
\caption{Online PSP}
\label{alg:pca}
\begin{algorithmic}
    \STATE {\bfseries input} $\{\x_t\}$; parameters $\gamma>0$ and $0<\eta<\tau$
    \STATE {\bfseries initialize} $\W\in\R^{k\times n}$ and $\M\in\mathbb{S}_{++}^k$
    \FOR{$t=1,2,\dots$}
        \REPEAT 
            \STATE $\z_t\gets \z_t+\gamma(\W\x_t-\M\z_t)$
        \UNTIL{convergence}
        \STATE $\W\gets\W+\eta(\z_t\x_t^\top-\W)$
        \STATE $\M\gets\M+\frac\eta\tau(\z_t\z_t^\top-\M)$
    \ENDFOR
\end{algorithmic}
\end{algorithm}
\vspace{10pt}

Algorithm \ref{alg:pca} can be mapped onto a multi-channel single-layer NN with $k$ point neurons and Hebbian plasticity. The neural dynamics are assumed to operate on a fast timescale and equilibrate before the synapses are updated. The synaptic updates to the feedforward weights $\W$ are naturally viewed as a combination of Hebbian and homeostatic plasticity. Since the first term in the synaptic update for the recurrent connections $-\M$ is \textit{inversely} proportional to product of the pre- and postsynaptic activities, the update is often referred to as \textit{anti-Hebbian}. Variations of the similarity matching objectives were used as starting points to derive Hebbian NNs for performing a number of unsupervised dimensionality reduction tasks that model aspects of early sensory processing \citep{pehlevan2019neuroscience}; however, these NNs cannot account for multi-compartment neurons and non-Hebbian forms of plasticity prevalent throughout the brain \citep{magee2020synaptic}.

\section{Multi-compartmental neurons and non-Hebbian plasticity}\label{sec:non-Hebbian}

Most neurons in the brain have multi-compartmental structures and learn via non-Hebbian synaptic plasticity rules. For example, pyramidal neurons, which are the main excitatory neurons of the cortex, receive feedforward excitatory inputs (e.g., lower level sensory inputs) via their proximal dendrites and feedback excitatory inputs (e.g., from farther up the cortical hierarchy) via dendrites that extend from the distal apical tuft \citep{steward1976cells,cauller1998backward,megias2001total,petreanu2009subcellular}, Figure \ref{fig:pca} (right). These inputs are integrated in at least 2 electronically segregated dendritic compartments---a proximal compartment near the soma of the pyramidal neuron and a distal compartment in the apical tuft \citep{magee1998dendritic,stuart1998determinants,harnett2013potassium,harnett2015distribution}. 
Integrated proximal feedforward inputs are the main drivers of the pyramidal neuron sodium action potential outputs \citep{turner1991site,stuart1994active}, while integrated distal feedback inputs generate calcium plateau potentials that are effective drivers of synaptic plasticity in the proximal dendrites \citep{schiller1997calcium,larkum1999new,golding2002dendritic,sjostrom2006cooperative,bittner2015}. 

There are a number of existing consequential models of both individual pyramidal neurons and other multi-compartmental neurons \citep{mel1994information,kording2000learning,poirazi2003pyramidal,poirazi2009information,hendrickson2011capabilities,urbanczik2014learning,guerguiev2017towards,sacramento2018dendritic,haga2018dendritic,richards2019dendritic,tzilivaki2019challenging,jones2021might,milstein2020bidirectional,larkum2022dendrites,mikulasch2023error,makarov2023dendrites}. These models provide detailed biophysical descriptions of the neural dynamics and non-Hebbian synaptic plasticity and, through numerical simulation, demonstrate computational capabilities of the pyramidal neuron and cortical circuits. However, these models do not provide a \textit{normative} framework for understanding multi-compartmental neurons and non-Hebbian forms of plasticity.


\section{CCA and symmetric generalized eigenvalue problems}\label{sec:gpsp}

To develop a normative framework, we first propose a class of computational objectives. Many linear versions of behaviorally relevant learning tasks can be formulated as symmetric generalized eigenvalue problems. Before stating the general problem, we first present the special case of CCA in the context of a pyramidal neuron.

Consider a pyramidal neuron that receives inputs from two upstream populations of neurons, whose activities at time $t$ are encoded as the components of the column vectors $\x_t\in\R^{n_x}$ and $\y_t\in\R^{n_y}$, Figure \ref{fig:pca} (right). One hypothesis is that the goal of the pyramidal neuron is to learn associations between these high-dimensional data streams. What should the objective be? A relevant associative learning objective is CCA \citep{hotelling1936relations}, which identifies subspaces of the input data streams such that the corresponding projections of the inputs are maximally correlated. 
In one-dimension, the objective is to find $n_x$-dimensional and $n_y$-dimensional row vectors $\w_x$ and $\w_y$ that maximize the covariance $\langle(\w_x\x_t)(\w_y\y_t)\rangle$ subject to the constraint $\langle(\w_x\x_t)^2\rangle=\langle(\w_y\y_t)^2\rangle=1$.

CCA is a special case of the symmetric generalized eigenvalue problem
\begin{align}\label{eq:gen_ev}
    \A\v=\lambda\B\v,&&\A:=\langle\bxi_t\bxi_t^\top\rangle,&&\B:=\langle\B_t\rangle,
\end{align}
when $\v^\top=[\w_x,\w_y]$ and
\begin{align}\label{eq:cca}
    \bxi_t=\begin{bmatrix}\x_t\\ \y_t\end{bmatrix},&&\B_t=\begin{bmatrix}\x_t\x_t^\top & \\ & \y_t\y_t^\top\end{bmatrix}.
\end{align}
We consider the class of symmetric generalized eigenvalue problems of the form in equation \eqref{eq:gen_ev}, where the pair $(\bxi_t,\B_t)\in\R^n\times\mathbb{S}_+^n$ is a function of the NN inputs; see Table \ref{tab:gevs} in section \ref{sec:nns} for specific examples.
Given such a symmetric generalized eigenvalue problem and $1\le k<n$, we refer to the projection of the vector $\bxi_t$ onto the $k$-dimensional subspace spanned by the top $k$ eigenvectors as the \textit{generalized principal subspace projection (GPSP)} of $(\bxi_t,\B_t)$.
Our goal is to derive an online multi-channel GPSP algorithm that maps onto a NN with \textit{local} learning rules, i.e., the synaptic updates only depend on variables that are represented in the pre- and postsynaptic neurons as well as globally broadcast variables. There are existing online CCA and GPSP algorithms \citep{arora2017stochastic,bhatia2018gen,pehlevan2020neurons,meng2021online,gemp2023generalized}; however, these algorithms cannot be mapped onto NNs with local learning rules and/or only find the top 1-dimensional projection.

\section{An online GPSP algorithm}\label{sec:derivation}

Here, we derive an online GPSP algorithm and, in the next section, we show that for many relevant examples the algorithm maps onto a NN with multi-compartmental neurons and local non-Hebbian learning rules. The reader who is not interested in the derivation can skip to the end of this section where we state our algorithm (Algorithm~\ref{alg:online}).

\subsection{Similarity matching objective}

At each time step $t$, let $\bzeta_t\in\R^k$ denote the GPSP of $(\bxi_t,\B_t)$. 
A useful observation is that $\bzeta_t$ is equal to the PSP of the normalized data $\sqrt{\B^\dag}\bxi_t$, where $\B^\dag$ is the Moore-Penrose inverse of $\B$---to see this, substitute in for $\A$ and $\v$ in equation \eqref{eq:gen_ev} with $\sqrt{\B^\dag}\A\sqrt{\B^\dag}$ and $\sqrt{\B}\v$, respectively. 
Therefore, we can substitute $\sqrt{\B^\dag}\bxi_t$ and $\bzeta_t$ in for $\x_t$ and $\z_t$, respectively, in the similarity matching objective \eqref{eq:objective_psp} for PSP to obtain the GPSP objective
\begin{align}\label{eq:objective}
    \min_{\bzeta_1,\dots,\bzeta_T\in\R^k}\frac{1}{T^2}\sum_{t=1}^T\sum_{t'=1}^T\left(\bxi_t^\top\B^\dag\bxi_{t'}-\bzeta_t^\top\bzeta_{t'}\right)^2.
\end{align}
Every optimal solution $\bzeta_t$ of the objective \eqref{eq:objective} is a PSP of the input $\sqrt{\B^\dag}\bxi_t$, which is a GPSP of the input data $(\bxi_t,\B_t)$. When $\bxi_t=\x_t$, $\bzeta_t=\z_t$ and $\B=\I_n$, we recover the similarity matching objective \eqref{eq:objective_psp} from \citep{pehlevan2015hebbian}.

\subsection{Matrix substitutions}

The objective \eqref{eq:objective} does not readily lead to an \textit{online} GPSP algorithm. For example, direct optimization of the objective via gradient descent with respect to the output $\bzeta_t$ requires taking gradient steps that depend on the inputs $(\bxi_{t'},\B_{t'})$ from every time point $t'=1,\dots,T$. Rather, following the approach of \citet{pehlevan2018similarity}, we substitute in with dynamic matrix variables to obtain a minimax algorithm that can be solved in the online setting. These matrix variables will correspond to feedforward and lateral synaptic weight matrices in the NN implementations.

For the cross term in equation \eqref{eq:objective}, we introduce the synaptic weight matrix $\W$ by substituting in with the Legendre transform
\begin{align*}
    -\frac{1}{T}\sum_{t=1}^T\bzeta_t^\top\left[\frac1T\sum_{t'=1}^T\bzeta_{t'}\bxi_{t'}^\top\B^\dag\right]\bxi_t&=\min_{\W\in\R^{k\times n}}\frac1T\sum_{t=1}^T\left[-2\bzeta_t^\top\W\bxi_t+\tr(\W\B_t\W^\top)\right].
\end{align*}
Differentiating the right-hand-side of the equality with respect to $\W$, setting the derivative to zero and solving for $\W$, we see that the optimum is achieved at $\frac1T\sum_{t'=1}^T\bzeta_{t'}\bxi_{t'}^\top\B^\dag$. To account for the quartic term in \eqref{eq:objective}, we introduce the synaptic weight matrix $\M$ by substituting in with Legendre transform
\begin{align*}
    \frac1T\sum_{t=1}^T\bzeta_t^\top\left[\frac1T\sum_{t'=1}^T\bzeta_{t'}\bzeta_{t'}^\top\right]\bzeta_t&=\max_{\M\in\mathbb{S}_{++}^k}\frac1T\sum_{t=1}^T\left[2\bzeta_t^\top\M\bzeta_t-\tr(\M^2)\right],
\end{align*}
where $\mathbb{S}_{++}^k$ denotes the set of $k\times k$ positive definite matrices. Differentiating the right-hand-side of the equality with respect to $\M$, setting the derivative to zero and solving for $\M$, we see that the optimum is achieved at $\frac1T\sum_{t'=1}^T\bzeta_{t'}\bzeta_{t'}^\top$. Substituting the Legendre transformations into the objective \eqref{eq:objective}, interchanging the order of optimization\footnote{Changing the order of optimization in this problem does not affect the solution due to the saddle point property, see \citep[section 5.4]{boyd2004convex}.} and dropping terms that do not depend on $\bzeta_t$, we arrive at the minimax objective:
\begin{align}\label{eq:minimax}
    \min_{\W\in\R^{k\times n}}\max_{\M\in\mathbb{S}_{++}^k}\frac1T\sum_{t=1}^T\min_{\bzeta_t\in\R^k}\ell(\W,\M,\bzeta_t,\bxi_t,\B_t),
\end{align} 
where
\begin{align}\label{eq:ell}
    \ell(\W,\M,\bzeta_t,\bxi_t,\B_t):=2\tr(\W\B_t\W^\top)-\tr(\M^2)-4\bzeta_t^\top\W\bxi_t+2\bzeta_t^\top\M\bzeta_t.
\end{align}
As a result of introducing the matrix variables, $\W$ and $\M$, we have transformed the minimization problem \eqref{eq:objective} into the minimax objective \eqref{eq:minimax}. This objective has the desirable property that for fixed $\W$ and $\M$, the optimal output $\bzeta_t$ at time step $t$ only depends on the input $\bxi_t$ at time step $t$. 

\subsection{Online algorithm}

To derive an online algorithm, we assume there is a separation of timescales between the minimization over the vectors $\bzeta_t$, which will correspond to neural activities, and the optimization of the matrices $\W$ and $\M$, which will correspond to synaptic weights.
At each time step $t$, we minimize $\ell(\W,\M,\bzeta_t,\bxi_t,\B_t)$ with respect to $\bzeta_t$ by running gradient descent steps until convergence
\begin{align}
    \label{eq:neural_dynamics}
    \bzeta_t\gets\bzeta_t+\gamma(\W\bxi_t-\M\bzeta_t)\qquad\Rightarrow\qquad\bzeta_t=\M^{-1}\W\bxi_t.
\end{align}
After $\bzeta_t$ equilibrates, we optimize $\langle\ell(\W,\M,\bzeta_t,\bxi_t,\B_t)\rangle$ with respect to the matrix variables by taking a stochastic gradient descent-ascent step in $\W$ and $\M$:
\begin{align}\label{eq:deltaWM}
    \W\gets\W+2\eta(\bzeta_t\bxi_t^\top-\W\B_t),&&\M\gets\M+\frac\eta\tau(\bzeta_t\bzeta_t^\top-\M).
\end{align}
Here $\eta>0$ is the step size for the stochastic gradient descent steps in $\W$ and $\tau>0$ denotes the ratio between the learning rate for $\W$ and the learning rate for $\M$. This yields our online GPSP algorithm, Algorithm~\ref{alg:online}.

\vspace{10pt}
\begin{algorithm}[H]
\caption{Online GPSP}
\label{alg:online}
\begin{algorithmic}
    \STATE {\bfseries input} $\{(\bxi_t,\B_t)\}$; parameters $\gamma>0$ and $0<\eta<\tau$
    \STATE {\bfseries initialize} $\W\in\R^{k\times n}$ and $\M\in\mathbb{S}_{++}^k$
    \FOR{$t=1,2,\dots$}
        \REPEAT 
            \STATE $\bzeta_t\gets \bzeta_t+\gamma(\W\bxi_t-\M\bzeta_t)$
        \UNTIL{convergence}
        \STATE $\W\gets\W+2\eta(\bzeta_t\bxi_t^\top-\W\B_t)$
        \STATE $\M\gets\M+\frac\eta\tau(\bzeta_t\bzeta_t^\top-\M)$
    \ENDFOR
\end{algorithmic}
\end{algorithm}
\vspace{10pt}

There are a few points worth noting:
\begin{itemize}
    \item Algorithm~\ref{alg:online} reduces to Algorithm~\ref{alg:pca} when $\bxi_t=\x_t$, $\bzeta_t=\z_t$ and $\B_t=\I_n$ for all $t$. 
    \item Since $\ell$ is nonconvex-concave in $\W$ and $\M$, the minimization over $\W$ cannot be interchanged with the maximization over $\M$ in equation \eqref{eq:minimax}. Therefore, to ensure convergence of the synaptic weights, the $\M$ updates need to be sufficiently fast relative to the $\W$ updates, i.e., $\tau>0$ needs to be sufficiently small.
    \item Since the symmetric generalized eigenvalue problem is defined in terms of the averages $\A:=\langle\bxi_t\bxi_t^\top\rangle$ and $\B:=\langle\B_t\rangle$ and the synaptic update rules are in terms of $(\bxi_t,\B_t)$, which are functions on the NN inputs, Algorithm~\ref{alg:online} establishes a precise relationship between the symmetric generalized eigenvalue problem and the synaptic learning rules via the variables $(\bxi_t,\B_t)$.
\end{itemize}
In general, the biological plausibility and biological interpretation of Algorithm \ref{alg:online} depends on the specific form of $\bxi_t$ and $\B_t$.

\section{Examples of NNs for GPSP}\label{sec:nns}

\begin{table}
    \centering
    \begin{tabular}{l c c c}
        Learning task & $\bxi_t$ & $\B_t$ & \# of compartments \\ [5pt] \hline & \\ 
        PCA & $\x_t$ & $\I_n$ & 1 \\ [20pt]
        CCA & $\begin{bmatrix}\x_t\\ \y_t\end{bmatrix}$ & $\begin{bmatrix}\x_t\x_t^\top & \\ & \y_t\y_t^\top \end{bmatrix}$ & 3 \\ [20pt] 
        SFA & $\x_t+\x_{t-1}$ & $\x_t\x_t^\top$ & 2 \\ [20pt]
        ICA (FOBI) & $\x_t$ & $\|\C_X^{-1/2}\x_t\|^2\x_t\x_t^\top$ & 2 \\ [20pt]
        cPCA* & $\delta_t\x_t$ & $(1-\delta_t)\x_t\x_t^\top$ & 2 \\ [5pt]
    \end{tabular}
    \caption{
    A list of learning tasks with symmetric generalized eigenvalue problem formulations that can be solved with NNs derived using our framework. 
    }
    \label{tab:gevs}
\end{table}

We consider several biologically relevant symmetric generalized eigenvalue problems that can be solved using Algorithm~\ref{alg:online}---for different choices of the vector $\bxi_t$ and matrix $\B_t$, Table \ref{tab:gevs}. For each symmetric generalized eigenvalue problem, we map its online algorithm onto a NN with multi-compartmental neurons and non-Hebbian learning rules, Figure \ref{fig:nns}.

\subsection{Canonical Correlation Analysis (CCA)}

\begin{figure}
    \centering
    \includegraphics[width=\textwidth]{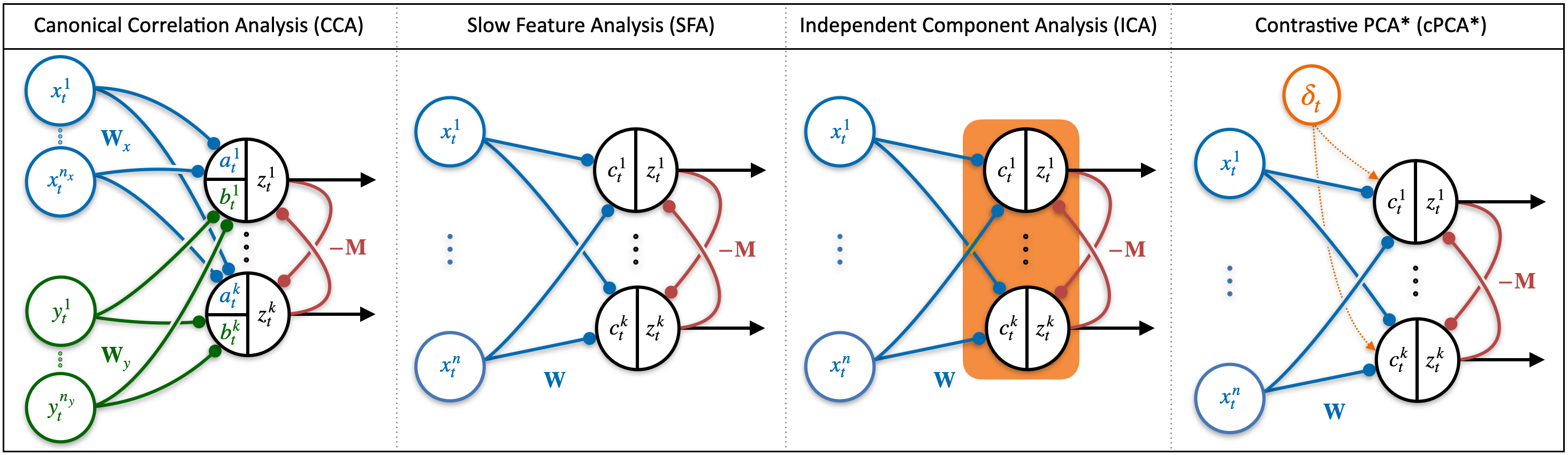}
    \caption{NNs for GPSP. In each NN, circles denote neurons and solid lines with circles at the ends are synapses. The synaptic updates depend on the variables inside the circles, which are encoded in different compartments of the neuron, and globally broadcast variables. 
    For ICA, the orange box denotes the presence of a scalar variable (specifically, $\|\z_t\|^2$) that is available to each output neuron. For cPCA*, the scalar $\delta_t$ is an indicator variable that gates the output of the NN and is available to each output neuron.
    }
    \label{fig:nns}
\end{figure}

As discussed in section \ref{sec:gpsp}, CCA may serve as a useful objective for understanding computation in pyramidal cells and cortical circuits.
Using our approach, we derived an online CCA algorithm that maps onto a NN with multi-compartmental neurons and non-Hebbian plasticity \citep{lipshutz2021biologically}. 
Substituting the expressions for $(\bxi_t,\B_t)$ from equation \eqref{eq:cca} into Algorithm~\ref{alg:online} results in an online algorithm that maps onto a single-layer NN, Figure \ref{fig:nns} (far left).

At each time step $t$, the NN receives inputs $\x_t$ and $\y_t$. The inputs are projected onto the feedforward synaptic weights $\W_x$ and $\W_y$, which combine to form the feedforward weight matrix $\W:=[\W_x,\W_y]$, to generate dendritic currents $\a_t=\W_x\x_t$ and $\b_t=\W_y\y_t$ that are stored in separate dendritic compartments. The output of the neurons $\z_t=\bzeta_t$, which is represented in a third compartment, is computed by running the recurrent neural dynamics:
\begin{align*}
    \z_t\gets\z_t+\gamma(\a_t+\b_t-\M\z_t)\quad\Rightarrow\quad\z_t=\M^{-1}(\a_t+\b_t).
\end{align*}
After the neural dynamic equilibrate, the synaptic weights are updated. The feedforward synaptic weight updates are given by
\begin{align*}
    \W_x\gets\W_x+2\eta(\z_t-\a_t)\x_t^\top,&&\W_y\gets\W_y+2\eta(\z_t-\b_t)\y_t^\top.
\end{align*}
Since the components of the vectors $\a_t$, $\b_t$ and $\z_t$ are represented in the postsynaptic neurons, the synaptic updates are local, but non-Hebbian. The lateral recurrent synaptic weight updates are as in Algorithm \ref{alg:online} with $\bzeta_t=\z_t$. 

The NN is consistent with certain aspects of experimentally observed physiology and anatomy of pyramidal neurons and cortical circuits. Each neuron includes 2 dendritic compartments that separately integrate the inputs $\x_t$ and $\y_t$. Rearranging the formula for the equilibrium neural outputs $\z_t$ of the NN, we see that $\b_t=\M\z_t-\a_t$ and so we can rewrite the proximal synaptic updates as
\begin{align*}
    \W_y\gets\W_y+2\eta(\a_t-[\M-\I_k]\z_t)\y_t^\top.
\end{align*}
From this formulation of the synaptic update, we can interpret the difference between the distal currents $\a_t$ and the recurrent lateral feedback $-[\M-\I_k]\z_t$ as the calcium plateau potential that drives non-Hebbian plasticity in the proximal synapses, which is consistent with experimental observations that distal currents generate calcium plateau potentials that drive plasticity in the proximal synapses and these plateaus are mediated by inhibitory inputs \citep{schiller1997calcium,larkum1999new,golding2002dendritic,sjostrom2006cooperative,bittner2015}. 

The NN for CCA includes direct lateral connections between the pyramidal cells; however, in cortical circuits, lateral communication is typically mediated by local interneurons. By modifying the starting CCA objective to include an output whitening constraint, we can derive an algorithm that faithfully maps onto the wiring diagram of a cortical microcircuit consisting of both pyramidal neurons and interneurons \citep{lipshutz2021biologically}. 

Finally, the output of the neurons in this circuit is symmetric in the integrated currents $\a_t$ and $\b_t$; however, experimental evidence suggests that the feedforward proximal inputs and feedback distal inputs are integrated asymmetrically by the pyramidal neuron \citep{turner1991site,stuart1994active,schiller1997calcium,larkum1999new,golding2002dendritic,sjostrom2006cooperative,bittner2015}, which is in contrast to our model that treats the inputs symmetrically. In \citep{golkar2020simple,golkar2020biologically}, we derived an online algorithm for CCA when interpreted as a \textit{supervised} learning task, where the feedforward inputs are feature vectors and the feedback inputs are supervisory signals. In this case, the output of the circuit is exclusively driven by the feedforward inputs.

\subsection{Slow Feature Analysis (SFA)}

Brains are adept at learning meaningful latent representations from noisy, high-dimensional data. Often, the relevant features in the environment (e.g., objects) change slowly compared with noisy sensory data, so temporal slowness has been proposed as a computational principle for identifying latent features \citep{foldiak1991learning,mitchison1991removing,wiskott2002slow}. A popular approach for extracting slow features, introduced by \citet{wiskott2002slow}, is called SFA. SFA is an unsupervised learning algorithm that extracts the slowest projection, in terms of discrete time derivative, from a nonlinear expansion of the input signal. When trained on natural images, SFA learns features that resemble properties of complex cells in the primary visual cortex \citep{berkes2005slow}. Further, when trained on a simulated visual stream, a hierarchical version of SFA learns representations of orientation and space similar to those encoded in the hippocampus \citep{franzius2007slowness}. Together, these observations suggest that the cortex may use temporal slowness as a computational principal in representation learning.

The projection of the nonlinear expansion can be formulated as a generalized eigenvalue problem of the form \eqref{eq:gen_ev} with $\bxi_t=\x_t+\x_{t-1}$ and $\B_t=\x_t\x_t^\top$. 
Substituting into Algorithm~\ref{alg:online} yields an online algorithm that maps onto a single-layer NN \citep{lipshutz2020biologically}, Figure \ref{fig:nns} (middle left).
At each time step $t$, the NN receives inputs $\x_t$, which are projected onto the weight matrix $\W$ to generate dendritic currents $\bfc_t=\W\x_t$. The dendritic currents $\bfc_t$ are stored in a separate compartment from the NN outputs $\z_t$, so each neuron consists of 2 compartments. Letting $\bzeta_t=\z_t+\z_{t-1}$ and assuming the weights do not significantly change between time steps $t-1$ and $t$, we can rewrite the neural dynamics in Algorithm \ref{alg:online} as
\begin{align*}
    \z_t\gets\z_t+\gamma(\bfc_t-\M\z_t)\quad\Rightarrow\quad\z_t=\M^{-1}\bfc_t.
\end{align*}
Substituting in with the definitions of $\bxi_t$, $\B_t$, $\bfc_t$ and $\bzeta_t$, the feedforward synaptic weight updates from Algorithm \ref{alg:online} are given by $\Delta\W=\eta((\z_t+\z_{t-1})(\x_t+\x_{t-1})^\top-\bfc_t\x_t^\top)$. The synaptic update depends on low frequency signals in both the pre- and postsynaptic neurons; however, dendrites are more likely to store low frequency signals than axons. 

If the input time series is stationary and reversible in time (i.e., $\langle\x_t\x_{t-1}^\top\rangle=\langle\x_{t-1}\x_t^\top\rangle$), we can rewrite the synaptic updates so they only depend on low frequency signals in the postsynaptic neurons and 
\begin{align*}
    \W&\gets\W+2\eta\left(2\z_t+2\z_{t-1}-\bfc_t\right)\x_t^\top.
\end{align*}
Empirically, this modification extracts slow signals even when the time series is not reversible \citep{lipshutz2020biologically}.
Assuming that the postsynaptic neurons represent their low-pass filtered activities $\z_t+\z_{t-1}$ as well as their dendritic currents $\bfc_t$, the learning rules only depend on variables that are available in the pre- and postsynaptic neurons and so the synaptic updates are local, but non-Hebbian.
Finally, the first term in the synaptic weights update resembles the first term in the following local ``trace rule'' proposed by \citet{foldiak1991learning} for learning temporally invariant features:
\begin{align*}
    \W_\text{trace}\gets\W_\text{trace}+\eta\left((\z_t+\z_{t-1})\x_t^\top-\text{diag}(\z_t+\z_{t-1})\W_\text{trace}\right).
\end{align*}
Therefore, this normative NN model establishes a relationship between a computational objective for SFA and a variant of the proposed trace rule.

\subsection{Independent Components Analysis (ICA)}\label{sec:ICA}

Efficient coding theories of sensory processing posit that early sensory layers transform their inputs to reduce redundancy \citep{barlow1989unsupervised,atick1992could}. ICA is a statistical method for reducing redundancy by factorizing the sensory inputs into ``independent components'' and can explain edge detector neurons in area V1 of the visual cortex \citep{bell1996edges,bell1997independent} and receptive fields of cochlear nerve fibers in the auditory system 
\citep{lewicki2002efficient}.

ICA assumes a generative model $\x_t=\A\s_t$, where $\s_t$ is the $n$-dimensional source vectors with independent components, $\A$ is a $n\times n$ mixing matrix and $\x_t$ is the $n$-dimensional mixture vector. One method for solving ICA, called Fourth-Order Blind Identification (FOBI) \citep{cardoso1989source}, assumes the components of the sources have distinct kurtosis (i.e., fourth-order moments). FOBI can be solved in three steps: (i) whitening the mixture vector: $\h_t=\C_X^{-1/2}\x_t$, (ii) weighting the whitened mixtures by their norms: $\y_t=\|\h_t\|\h_t$ and (iii) projecting the whitened mixtures $\h_t$ onto the principal components of $\y_t$. Remarkably, these three steps can be combined and expressed as a single symmetric generalized eigenvalue problem of the form \eqref{eq:gen_ev} with $\bxi_t=\x_t$ and $\B_t=\|\h_t\|^2\x_t\x_t^\top$, so we can apply our framework. 
Substituting into Algorithm~\ref{alg:online} results in an online algorithm that maps onto a single-layer NN with multi-compartmental neurons and local non-Hebbian learning rules \citep{bahroun2021normative}, Figure \ref{fig:nns} (middle right).

At each time step $t$, the NN receives inputs $\x_t$, which are projected onto the feedforward synaptic weights $\W$ to generate dendritic currents $\bfc_t=\W\x_t$. The dendritic currents $\bfc_t$ are stored in a separate compartment from the NN outputs $\z_t=\bzeta_t$, so each neuron consists of 2 compartments.
The neural dynamics in Algorithm \ref{alg:online} can be written as
\begin{align*}
    \z_t\gets\z_t+\gamma(\bfc-\M\z_t)\quad\Rightarrow\quad\z_t=\M^{-1}\bfc.
\end{align*}
Substituting the expression for $\B_t$ into the feedforward synaptic weight update in Algorithm \ref{alg:online} results in the update $\Delta\W=\eta(\z_t-\|\h_t\|^2\bfc_t)\x_t^\top$. As stated in step (iii) of FOBI, the outputs $\z_t$ are equal to the (full-rank) projection of the whitened inputs $\h_t$ onto the principal components of $\y_t$, which implies that $\z_t$ is an orthogonal transformation of the whitened inputs at each time $t$. Therefore, $\|\h_t\|=\|\z_t\|$ and we can rewrite the feedforward synaptic update as
\begin{align*}
    \W\gets\W+2\eta(\z_t-\|\z_t\|^2\bfc_t)\x_t^\top.
\end{align*}
Interestingly, the resulting synaptic learning rules are globally modulated by the total activity of the output neurons $\|\z_t\|^2$, which could be accounted for by biophysical quantities such as neuromodulators, extracellular calcium, local field potential, or nitric oxide.

\subsection{Contrastive Principal Component Analysis* (cPCA*)}

Sensory organs receive an immense amount of information per unit time, but much of it is of little relevance for behavior. A simple approach to process this high-dimensional input is to focus on a lower-dimensional subspace and ignore the directions that are less informative. PCA achieves this by discarding the directions with low variance. Such an approach is, however, inefficient in cases where the irrelevant directions are very noisy, thus having greater variance than the relevant ones. If we have access to representative samples  of variability in irrelevant directions (``negative samples''), we can achieve better efficiency by using a contrastive variant of PCA, as in~\cite{abid2018exploring}. Contrastive PCA (cPCA) finds the subspace of highest \emph{relevant} variance, associated with ``positive samples'', while minimizing the variance associated with irrelevant information, as inferred from negative samples.

In \citep{golkar2022cpca}, we consider a more robust cPCA method, which we refer to as cPCA* and derive an online algorithm with a neural implementation. Assume we have a sequence of centered inputs $(\x_1,\delta_1),\dots,(\x_T,\delta_T)\in\R^n\times\{0,1\}$. At each time $t$, the input $\x_t$ is a feature vector that is either a \textit{positive sample} or a \textit{negative sample}, so the positive and negative samples arrive via the same pathway. The scalar variable $\delta_t$ is equal to 1 (resp.\ 0) if the $\x_t$ is a positive (resp.\ negative) sample. 
The covariance matrices for the positive and negative samples are respectively given by $\C_{(+)}:=\langle\x_t\x_t^\top|\delta_t=1\rangle$ and $\C_{(-)}:=\langle\x_t\x_t^\top|\delta_t=0\rangle$.
The goal of cPCA* is to project the feature vectors $\x_t$ onto vectors $\v$ to maximize the ratio of $\v^\top\C_{(+)}\v$ and $\v^\top\C_{(-)}\v$, which corresponds to the symmetric generalized eigenvalue problem of the form \eqref{eq:gen_ev} with $\bxi_t:=\delta_t\x_t$ and $\B_t:=(1-\delta_t)\x_t\x_t^\top$. 
When the positive samples are measurements of signal + noise and the negative samples are measurements of nosie, the problem is closely related to linear discriminant analysis \citep{mclachlan2005discriminant} and joint decorrelation methods \citep{de2014joint}.
Substituting into Algorithm~\ref{alg:online} results in an online algorithm that maps onto a single-layer NN with multi-compartmental neurons and local non-Hebbian learning rules, Figure \ref{fig:nns} (far right). 

At each time step $t$, the NN receives inputs $\x_t$ and $\delta_t$. The inputs $\x_t$ are projected onto the feedforward weights $\W$ to generate dendritic currents $\bfc_t=\W\x_t$. The neural dynamics and synaptic updates for positive and negative samples are given by
\begin{table}[h!]
    \centering
    \begin{tabular}{ c c c}
        $\delta_t$ & fast neural dynamics & slow synaptic updates \\ [5pt] \hline & \\ 
        1 & $\z_t\gets\z_t+\gamma(\bfc_t-\M\z_t)$ & $\W\gets\W+2\eta\z_t\x_t^\top$ \\ [20pt]
        0 & $\z_t={\bf 0}$ & $\W\gets\W-2\eta\bfc_t\x_t^\top$ \\ [5pt]
    \end{tabular}
\end{table}

\noindent The scalar $\delta_t$ is naturally interpreted as indicated the presence or absences of a neuromodulator that gates the output $\z_t$ of the NN. Assuming the dendritic currents $\bfc_t$ and neural outputs $\z_t$ are represented in the postsynaptic neurons and $\delta_t$ is globally available, the synaptic updates are local and non-Hebbian. In both cases, the lateral synapses $-\M$ are updated according to Algorithm \ref{alg:online} with $\bzeta_t=\z_t$.

\section{Nonnegative Similarity Matching}\label{sec:nn_sm}

Most biological neurons perform nonlinear transformations and have nonnegative outputs. In addition, many interesting computations require nonlinear transformations. We can adapt the objective to account for both the nonlinear transformation and the nonnegativity of the neural outputs by imposing a nonnegativity constraint on the outputs $\bzeta_t$ in equation \eqref{eq:objective}, which results in the nonnegative similarity matching objective:
\begin{align}\label{eq:nnobjective}
    \min_{\bzeta_1,\dots,\bzeta_T\in\R_+^k}\frac{1}{T^2}\sum_{t=1}^T\sum_{t'=1}^T\left(\bxi_t^\top\B^\dag\bxi_{t'}-\bzeta_t^\top\bzeta_{t'}\right)^2,
\end{align}
where $\R_+^k$ denotes the nonnegative orthant in $\R^k$. Imposing the nonnegative constraint transforms the problem from a spectral matrix factorization problem to a nonnegative matrix factorization problem. For the special case that $\B=\I_n$, \citet{pehlevan2014hebbian,bahroun2017online,sengupta2018manifold,qin2021contrastive} explored the relationship between the objective \eqref{eq:nnobjective} and clustering, sparse representation learning, manifold tiling and supervised learning. However, aside from a specific generative model (see Sec.~\ref{sec:nica} below), it is not clear how to interpret objective \eqref{eq:nnobjective} when $\B$ is not the identity matrix.

\subsection{An online algorithm}

We can derive an online algorithm for solving the objective \eqref{eq:nnobjective} following the same steps as in Sec.~\ref{sec:derivation} for deriving an online GPSP algorithm. First, we perform the same matrix substitutions to arrive at the minimax problem
\begin{align}\label{eq:minimax_nn}
    \min_{\W\in\R^{k\times n}}\max_{\M\in\mathbb{S}_{++}^k}\min_{\bzeta_1,\dots,\bzeta_T\in\R_+^k}\langle\ell(\W,\M,\bzeta_t,\bxi_t,\B_t)\rangle,
\end{align} 
where $\ell$ is defined as in equation \eqref{eq:ell}. To solve the minimax problem \eqref{eq:minimax_nn} in the online setting, at each time step $t$, we first minimize $\ell$ with respect to $\bzeta_t\in\R_+^k$ by taking \textit{projected} gradient steps until convergence:
\begin{align*}
    \bzeta_t\gets[\bzeta_t+\gamma(\W\bxi_t-\M\bzeta_t)]_+,
\end{align*}
where $[\cdot]_+$ denotes taking the nonnegative part elementwise. The minimization is still over a convex set; however, unlike the GPSP setting, we do not have a closed-form expression for the output $\bzeta_t$. After $\bzeta_t$ converges, we update the matrices $\W$ and $\M$ by taking a stochastic gradient decsent-ascent step, which results in the exact same updates as in equation \eqref{eq:deltaWM}.

\subsection{Nonnegative Independent Component Analysis (NICA)}
\label{sec:nica}

As discussed in Sec.~\ref{sec:ICA}, ICA is a statistical method for factorizing sensory inputs into independent components. A special case is called Nonnegative ICA (NICA), which assumes a generative model in which the mixture of stimuli is a linear combination of uncorrelated, nonnegative sources; i.e., $\x_t=\A\s_t$, where $\s_t$ denotes the nonnegative vector of source intensities, $\A$ is a mixing matrix and $\x_t$ denotes the vector of mixed stimuli. The goal of NICA is to infer the nonnegative source vectors $\s_t$ from the mixture vectors $\x_t$. Both the linear additivity of stimuli and nonnegativity of the sources are reasonable assumptions in biological applications. For example, in olfaction, concentrations of odorants are both additive and nonnegative.

While NICA cannot be expressed as a GPSP problem, it can be solved using the nonnegative similarity matching framework. \citet{pehlevan2017blind} solved NICA with an online algorithm that can be implemented in a 2-layer network with point neurons and Hebbian/anti-Hebbian learning rules, where each layer is derived from a separate objective function. The 2 objective functions can be combined into a single nonnegative similarity matching objective of the form \eqref{eq:nnobjective} with $\bxi_t=\x_t$ and $\B_t=(\x_t-\langle\x_t\rangle)(\x_t-\langle\x_t\rangle)^\top$. Starting from the nonnegative similarity matching objective, we derived an online algorithm for solving NICA that maps onto a single-layer network with multi-compartmental neurons and non-Hebbian plasticity \citep{lipshutz2022single}. 

At each time step $t$ the NN receives inputs $\x_t$ which is projected onto the feedforward weights to generate the dendritic current $\bfc_t$, which is stored in a separate compartment from the NN outputs $\z_t=\bzeta_t$. The fast neural dynamics are given by
\begin{align*}
    \z_t\gets[\z_t+\gamma(\bfc_t-\M\z_t)]_+.
\end{align*}
After the neural dynamics equilibrate, the feedforward synaptic weights are updated according to the learning rule
\begin{align*}
    \W&\gets\W+2\eta\left(\z_t\x_t^\top-(\bfc_t-\bar\bfc_t)(\x_t-\bar\x_t)^\top\right),
\end{align*}
where we have replaced $\W\langle\x_t\rangle$ (resp.\ $\langle\x_t\rangle$) with the running average $\bar\bfc_t$ (resp.\ $\bar\x_t$) of the dendritic current (resp.\ inputs), which could be physically represented as local ion concentrations at the synapses. 

\section{Discussion}\label{sec:discussion}

In this work, we proposed an extension of the similarity matching objective to include a broad class of symmetric generalized eigenvalue problems. Starting from this objective, we derived an online algorithm and showed that for several examples, the algorithm maps onto a NN with multi-compartmental neurons and local, non-Hebbian learning rules. Furthermore, we proposed an modification of our framework to solve a broad class of nonnegative matrix factorization problems and we mapped a specific example onto a NN with multi-compartmental neurons, local learning rules and rectified outputs. 

Our framework establishes a precise relationship between synaptic learning rules and computational objectives. In particular, the synaptic learning rules in Algorithm \ref{alg:online} are related to the symmetric generalized eigenvalue problem \eqref{eq:gen_ev} via the variables $\bxi_t$ and $\B_t$. Therefore, given a symmetric generalized eigenvalue problem of the form \eqref{eq:gen_ev}, one can predict the synaptic learning rules for the NN. Conversely, given synaptic learning rules of the form in Algorithm \ref{alg:online}, one can predict the computational objective for the NN. We believe this unified framework for relating non-Hebbian synaptic learning rules to computational objectives will be useful for understanding forms of non-Hebbian plasticity found throughout the brain.

Finally, in addition to the framework presented here, there are other normative approaches for deriving online algorithms that map onto NNs with multi-compartmental neurons and solve symmetric generalized eigenvalue problems and other related problems \citep{golkar2020simple,golkar2022constrained,friedrich2021neural,lipshutz2022linear,duong2023adaptive}.


\subsection*{Acknowledgements}

We thank Pierre-\'Etienne Fiquet for helpful feedback on an earlier draft of this work.

\bibliographystyle{unsrtnat}
\bibliography{biblio.bib}

\begin{thebibliography}{115}
\providecommand{\natexlab}[1]{#1}
\providecommand{\url}[1]{\texttt{#1}}
\expandafter\ifx\csname urlstyle\endcsname\relax
  \providecommand{\doi}[1]{doi: #1}\else
  \providecommand{\doi}{doi: \begingroup \urlstyle{rm}\Url}\fi

\bibitem[Attneave(1954)]{attneave1954some}
Fred Attneave.
\newblock Some informational aspects of visual perception.
\newblock \emph{Psychological review}, 61\penalty0 (3):\penalty0 183, 1954.

\bibitem[Barlow(1961)]{barlow1961possible}
Horace~B Barlow.
\newblock Possible principles underlying the transformation of sensory
  messages.
\newblock \emph{Sensory Communication}, 1\penalty0 (1):\penalty0 217--233,
  1961.

\bibitem[Srinivasan et~al.(1982)Srinivasan, Laughlin, and
  Dubs]{srinivasan1982predictive}
Mandyam~Veerambudi Srinivasan, Simon~Barry Laughlin, and Andreas Dubs.
\newblock Predictive coding: a fresh view of inhibition in the retina.
\newblock \emph{Proceedings of the Royal Society of London. Series B.
  Biological Sciences}, 216\penalty0 (1205):\penalty0 427--459, 1982.

\bibitem[Oja(1982)]{oja1982simplified}
Erkki Oja.
\newblock Simplified neuron model as a principal component analyzer.
\newblock \emph{Journal of Mathematical Biology}, 15:\penalty0 267--273, 1982.

\bibitem[Atick and Redlich(1992)]{atick1992does}
Joseph~J Atick and A~Norman Redlich.
\newblock What does the retina know about natural scenes?
\newblock \emph{Neural Computation}, 4\penalty0 (2):\penalty0 196--210, 1992.

\bibitem[van Hateren(1992)]{van1992theory}
Johannes~H van Hateren.
\newblock A theory of maximizing sensory information.
\newblock \emph{Biological Cybernetics}, 68\penalty0 (1):\penalty0 23--29,
  1992.

\bibitem[Olshausen and Field(1997)]{olshausen1997sparse}
Bruno~A Olshausen and David~J Field.
\newblock Sparse coding with an overcomplete basis set: A strategy employed by
  {V1}?
\newblock \emph{Vision Research}, 37\penalty0 (23):\penalty0 3311--3325, 1997.

\bibitem[Rao and Ballard(1999)]{rao1999predictive}
Rajesh~PN Rao and Dana~H Ballard.
\newblock Predictive coding in the visual cortex: a functional interpretation
  of some extra-classical receptive-field effects.
\newblock \emph{Nature Neuroscience}, 2\penalty0 (1):\penalty0 79--87, 1999.

\bibitem[Chen et~al.(2006)Chen, Hall, and Chklovskii]{chen2006wiring}
Beth~L Chen, David~H Hall, and Dmitri~B Chklovskii.
\newblock Wiring optimization can relate neuronal structure and function.
\newblock \emph{Proceedings of the National Academy of Sciences}, 103\penalty0
  (12):\penalty0 4723--4728, 2006.

\bibitem[Pehlevan et~al.(2015)Pehlevan, Hu, and
  Chklovskii]{pehlevan2015hebbian}
Cengiz Pehlevan, Tao Hu, and Dmitri~B Chklovskii.
\newblock A {H}ebbian/anti-{H}ebbian neural network for linear subspace
  learning: A derivation from multidimensional scaling of streaming data.
\newblock \emph{Neural Computation}, 27\penalty0 (7):\penalty0
  1461{\textendash}1495, 2015.

\bibitem[M{\l}ynarski and Hermundstad(2021)]{mlynarski2021efficient}
Wiktor~F M{\l}ynarski and Ann~M Hermundstad.
\newblock Efficient and adaptive sensory codes.
\newblock \emph{Nature Neuroscience}, 24\penalty0 (7):\penalty0 998--1009,
  2021.

\bibitem[Pearson(1901)]{pearson1901liii}
Karl Pearson.
\newblock {LIII}. on lines and planes of closest fit to systems of points in
  space.
\newblock \emph{The London, Edinburgh, and Dublin philosophical magazine and
  journal of science}, 2\penalty0 (11):\penalty0 559--572, 1901.

\bibitem[Hebb(1949)]{hebb1949organisation}
Donald~Olding Hebb.
\newblock \emph{The Organisation of Behaviour: A Neuropsychological Theory}.
\newblock Science Editions New York, 1949.

\bibitem[Bliss and Gardner-Medwin(1973)]{bliss1973longA}
Tim~VP Bliss and AR~Gardner-Medwin.
\newblock Long-lasting potentiation of synaptic transmission in the dentate
  area of the unanaesthetized rabbit following stimulation of the perforant
  path.
\newblock \emph{The Journal of Physiology}, 232\penalty0 (2):\penalty0 357,
  1973.

\bibitem[Bliss and L{\o}mo(1973)]{bliss1973longB}
Tim~VP Bliss and Terje L{\o}mo.
\newblock Long-lasting potentiation of synaptic transmission in the dentate
  area of the anaesthetized rabbit following stimulation of the perforant path.
\newblock \emph{The Journal of Physiology}, 232\penalty0 (2):\penalty0
  331--356, 1973.

\bibitem[Caporale and Dan(2008)]{caporale2008spike}
Natalia Caporale and Yang Dan.
\newblock Spike timing--dependent plasticity: a {H}ebbian learning rule.
\newblock \emph{Annual Review Neuroscience}, 31:\penalty0 25--46, 2008.

\bibitem[Oja and Karhunen(1983)]{oja1983analysis}
Erkki Oja and Juha Karhunen.
\newblock An analysis of convergence for a learning version of the subspace
  method.
\newblock \emph{Journal of Mathematical Analysis and Applications}, 91\penalty0
  (1):\penalty0 102--111, 1983.

\bibitem[Sanger(1989)]{sanger1989optimal}
Terence~D Sanger.
\newblock Optimal unsupervised learning in a single-layer linear feedforward
  neural network.
\newblock \emph{Neural Networks}, 2\penalty0 (6):\penalty0 459--473, 1989.

\bibitem[Leen et~al.(1990)Leen, Rudnick, and Hammerstrom]{leen1990hebbian}
Todd Leen, Mike Rudnick, and Dan Hammerstrom.
\newblock Hebbian feature discovery improves classifier efficiency.
\newblock In \emph{1990 IJCNN International Joint Conference on Neural
  Networks}, pages 51--56. IEEE, 1990.

\bibitem[Pehlevan and Chklovskii(2019)]{pehlevan2019neuroscience}
Cengiz Pehlevan and Dmitri~B Chklovskii.
\newblock Neuroscience-inspired online unsupervised learning algorithms:
  Artificial neural networks.
\newblock \emph{IEEE Signal Processing Magazine}, 36\penalty0 (6):\penalty0
  88--96, 2019.

\bibitem[Obeid et~al.(2019)Obeid, Ramambason, and Pehlevan]{deepSM}
Dina Obeid, Hugo Ramambason, and Cengiz Pehlevan.
\newblock Structured and deep similarity matching via structured and deep
  {H}ebbian networks.
\newblock In \emph{Advances in Neural Information Processing Systems}, 2019.

\bibitem[Te{\c{s}}ileanu et~al.(2022)Te{\c{s}}ileanu, Golkar, Nasiri, Sengupta,
  and Chklovskii]{tecsileanu2022neural}
Tiberiu Te{\c{s}}ileanu, Siavash Golkar, Samaneh Nasiri, Anirvan~M Sengupta,
  and Dmitri~B Chklovskii.
\newblock Neural circuits for dynamics-based segmentation of time series.
\newblock \emph{Neural Computation}, 34\penalty0 (4):\penalty0 891--938, 2022.

\bibitem[Genkin et~al.(2022)Genkin, Lipshutz, Golkar, Te{\c{s}}ileanu, and
  Chklovskii]{genkin2022transformation}
Alexander Genkin, David Lipshutz, Siavash Golkar, Tiberiu Te{\c{s}}ileanu, and
  Dmitri~B Chklovskii.
\newblock Biological learning of irreducible representations of commuting
  transformations.
\newblock \emph{Advances in Neural Information Processing Systems}, 35, 2022.

\bibitem[Pehlevan et~al.(2017{\natexlab{a}})Pehlevan, Genkin, and
  Chklovskii]{pehlevan2017clustering}
Cengiz Pehlevan, Alexander Genkin, and Dmitri~B Chklovskii.
\newblock A clustering neural network model of insect olfaction.
\newblock In \emph{2017 51st Asilomar Conference on Signals, Systems, and
  Computers}, page 593{\textendash}600. IEEE, IEEE, 2017{\natexlab{a}}.

\bibitem[Bahroun et~al.(2019)Bahroun, Chklovskii, and
  Sengupta]{bahroun2019similarity}
Yanis Bahroun, Dmitri Chklovskii, and Anirvan Sengupta.
\newblock A similarity-preserving network trained on transformed images
  recapitulates salient features of the fly motion detection circuit.
\newblock \emph{Advances in Neural Information Processing Systems}, 32, 2019.

\bibitem[Benna and Fusi(2021)]{benna2021place}
Marcus~K Benna and Stefano Fusi.
\newblock Place cells may simply be memory cells: Memory compression leads to
  spatial tuning and history dependence.
\newblock \emph{Proceedings of the National Academy of Sciences}, 118\penalty0
  (51):\penalty0 e2018422118, 2021.

\bibitem[Chapochnikov et~al.(2023)Chapochnikov, Pehlevan, and
  Chklovskii]{chapochnikov}
Nikolai~M Chapochnikov, Cengiz Pehlevan, and Dmitri~B Chklovskii.
\newblock Normative and mechanistic model of an adaptive circuit for efficient
  encoding and feature extraction.
\newblock \emph{Proceedings of the National Academy of Sciences}, 2023.
\newblock In press.

\bibitem[Qin et~al.(2023)Qin, Farashahi, Lipshutz, Sengupta, Chklovskii, and
  Pehlevan]{drift}
Shanshan Qin, Shiva Farashahi, David Lipshutz, Anirvan~M Sengupta, Dmitri~B
  Chklovskii, and Cengiz Pehlevan.
\newblock Coordinated drift of receptive fields in {H}ebbian/anti-{H}ebbian
  network models during noisy representation learning.
\newblock \emph{Nature Neuroscience}, 26:\penalty0 339--349, 2023.

\bibitem[Lipshutz et~al.(2023{\natexlab{a}})Lipshutz, Pehlevan, and
  Chklovskii]{lipshutz2022interneurons}
David Lipshutz, Cengiz Pehlevan, and Dmitri~B Chklovskii.
\newblock Interneurons accelerate learning dynamics in recurrent neural
  networks for statistical adaptation.
\newblock \emph{International Conference on Learning Representations},
  2023{\natexlab{a}}.

\bibitem[Magee and Grienberger(2020)]{magee2020synaptic}
Jeffrey~C Magee and Christine Grienberger.
\newblock Synaptic plasticity forms and functions.
\newblock \emph{Annual Review of Neuroscience}, 43:\penalty0 95--117, 2020.

\bibitem[Gidon et~al.(2020)Gidon, Zolnik, Fidzinski, Bolduan, Papoutsi,
  Poirazi, Holtkamp, Vida, and Larkum]{gidon2020dendritic}
Albert Gidon, Timothy~Adam Zolnik, Pawel Fidzinski, Felix Bolduan, Athanasia
  Papoutsi, Panayiota Poirazi, Martin Holtkamp, Imre Vida, and Matthew~Evan
  Larkum.
\newblock Dendritic action potentials and computation in human layer 2/3
  cortical neurons.
\newblock \emph{Science}, 367\penalty0 (6473):\penalty0 83--87, 2020.

\bibitem[Spruston(2008)]{spruston2008pyramidal}
Nelson Spruston.
\newblock Pyramidal neurons: dendritic structure and synaptic integration.
\newblock \emph{Nature Reviews Neuroscience}, 9\penalty0 (3):\penalty0
  206--221, 2008.

\bibitem[Takahashi and Magee(2009)]{takahashi2009pathway}
Hiroto Takahashi and Jeffrey~C Magee.
\newblock Pathway interactions and synaptic plasticity in the dendritic tuft
  regions of {CA1} pyramidal neurons.
\newblock \emph{Neuron}, 62\penalty0 (1):\penalty0 102--111, 2009.

\bibitem[Lipshutz et~al.(2021)Lipshutz, Bahroun, Golkar, Sengupta, and
  Chklovskii]{lipshutz2021biologically}
David Lipshutz, Yanis Bahroun, Siavash Golkar, Anirvan~M Sengupta, and Dmitri~B
  Chklovskii.
\newblock A biologically plausible neural network for multichannel canonical
  correlation analysis.
\newblock \emph{Neural Computation}, 33\penalty0 (9):\penalty0 2309--2352,
  2021.

\bibitem[Lipshutz et~al.(2020)Lipshutz, Windolf, Golkar, and
  Chklovskii]{lipshutz2020biologically}
David Lipshutz, Charles Windolf, Siavash Golkar, and Dmitri~B Chklovskii.
\newblock A biologically plausible neural network for slow feature analysis.
\newblock \emph{Advances in Neural Information Processing Systems},
  33:\penalty0 14986--14996, 2020.

\bibitem[Bahroun et~al.(2021)Bahroun, Chklovskii, and
  Sengupta]{bahroun2021normative}
Yanis Bahroun, Dmitri Chklovskii, and Anirvan Sengupta.
\newblock A normative and biologically plausible algorithm for independent
  component analysis.
\newblock \emph{Advances in Neural Information Processing Systems},
  34:\penalty0 7368--7384, 2021.

\bibitem[Golkar et~al.(2023)Golkar, Lipshutz, Tesileanu, and
  Chklovskii]{golkar2022cpca}
Siavash Golkar, David Lipshutz, Tiberiu Tesileanu, and Dmitri~B Chklovskii.
\newblock An online algorithm for contrastive principal component analysis.
\newblock \emph{IEEE International Conference on Acoustics, Speech and Signal
  Processing}, 2023.

\bibitem[Lipshutz et~al.(2022)Lipshutz, Pehlevan, and
  Chklovskii]{lipshutz2022single}
David Lipshutz, Cengiz Pehlevan, and Dmitri~B Chklovskii.
\newblock Biologically plausible single-layer networks for nonnegative
  independent component analysis.
\newblock \emph{Biological Cybernetics}, 116\penalty0 (5--6):\penalty0
  557--568, 2022.

\bibitem[Cunningham and Ghahramani(2015)]{cunningham2015linear}
John~P Cunningham and Zoubin Ghahramani.
\newblock Linear dimensionality reduction: Survey, insights, and
  generalizations.
\newblock \emph{The Journal of Machine Learning Research}, 16\penalty0
  (1):\penalty0 2859--2900, 2015.

\bibitem[Ghojogh et~al.(2019)Ghojogh, Karray, and
  Crowley]{ghojogh2019eigenvalue}
Benyamin Ghojogh, Fakhri Karray, and Mark Crowley.
\newblock Eigenvalue and generalized eigenvalue problems: Tutorial.
\newblock \emph{arXiv preprint arXiv:1903.11240}, 2019.

\bibitem[Atick and Redlich(1990)]{atick1990towards}
Joseph~J Atick and A~Norman Redlich.
\newblock Towards a theory of early visual processing.
\newblock \emph{Neural Computation}, 2\penalty0 (3):\penalty0 308--320, 1990.

\bibitem[Ganguli and Sompolinsky(2012)]{ganguli2012compressed}
Surya Ganguli and Haim Sompolinsky.
\newblock Compressed sensing, sparsity, and dimensionality in neuronal
  information processing and data analysis.
\newblock \emph{Annual Review of Neuroscience}, 35:\penalty0 485--508, 2012.

\bibitem[Hubel(1995)]{hubel1995eye}
David~H Hubel.
\newblock \emph{Eye, brain, and vision}.
\newblock Scientific American Library/Scientific American Books, 1995.

\bibitem[Yang(1995)]{yang1995projection}
Bin Yang.
\newblock Projection approximation subspace tracking.
\newblock \emph{IEEE Transactions on Signal processing}, 43\penalty0
  (1):\penalty0 95--107, 1995.

\bibitem[Duflo(2013)]{duflo2013random}
Marie Duflo.
\newblock \emph{Random iterative models}, volume~34.
\newblock Springer Science \& Business Media, 2013.

\bibitem[Chou and Wang(2020)]{chou2020ode}
Chi-Ning Chou and Mien~Brabeeba Wang.
\newblock Ode-inspired analysis for the biological version of {O}ja’s rule in
  solving streaming {PCA}.
\newblock In \emph{Conference on Learning Theory}, pages 1339--1343. PMLR,
  2020.

\bibitem[F\"oldiak(1989)]{foldiak1989adaptive}
P~F\"oldiak.
\newblock Adaptive network for optimal linear feature extraction.
\newblock In \emph{Proceedings of IEEE/INNS Int. Joint. Conf. Neural Networks},
  volume~1, pages 401--405, 1989.

\bibitem[Rubner and Tavan(1989)]{rubner1989self}
Jeanne Rubner and Paul Tavan.
\newblock A self-organizing network for principal-component analysis.
\newblock \emph{EPL (Europhysics Letters)}, 10\penalty0 (7):\penalty0 693,
  1989.

\bibitem[Rubner and Schulten(1990)]{rubner1990development}
Jeanette Rubner and Klaus Schulten.
\newblock Development of feature detectors by self-organization.
\newblock \emph{Biological Cybernetics}, 62\penalty0 (3):\penalty0 193--199,
  1990.

\bibitem[Leen(1991)]{leen1991learning}
Todd~K Leen.
\newblock Learning in linear feature-discovery networks.
\newblock In \emph{Adaptive Signal Processing}, volume 1565, pages 472--481.
  International Society for Optics and Photonics, 1991.

\bibitem[Pehlevan et~al.(2018)Pehlevan, Sengupta, and
  Chklovskii]{pehlevan2018similarity}
Cengiz Pehlevan, Anirvan~M Sengupta, and Dmitri~B Chklovskii.
\newblock Why do similarity matching objectives lead to
  {H}ebbian/anti-{H}ebbian networks?
\newblock \emph{Neural Computation}, 30\penalty0 (1):\penalty0
  84{\textendash}124, 2018.

\bibitem[Cox and Cox(2008)]{cox2008multidimensional}
Michael~AA Cox and Trevor~F Cox.
\newblock Multidimensional scaling.
\newblock In \emph{Handbook of Data Visualization}, pages 315--347. Springer,
  2008.

\bibitem[Steward and Scoville(1976)]{steward1976cells}
Oswald Steward and Sheila~A Scoville.
\newblock Cells of origin of entorhinal cortical afferents to the hippocampus
  and fascia dentata of the rat.
\newblock \emph{Journal of Comparative Neurology}, 169\penalty0 (3):\penalty0
  347--370, 1976.

\bibitem[Cauller et~al.(1998)Cauller, Clancy, and Connors]{cauller1998backward}
Lawrence~J Cauller, Barbara Clancy, and Barry~W Connors.
\newblock Backward cortical projections to primary somatosensory cortex in rats
  extend long horizontal axons in layer {I}.
\newblock \emph{Journal of Comparative Neurology}, 390\penalty0 (2):\penalty0
  297--310, 1998.

\bibitem[Meg{\i}as et~al.(2001)Meg{\i}as, Emri, Freund, and
  Gulyas]{megias2001total}
M~Meg{\i}as, ZS~Emri, TF~Freund, and AI~Gulyas.
\newblock Total number and distribution of inhibitory and excitatory synapses
  on hippocampal {CA1} pyramidal cells.
\newblock \emph{Neuroscience}, 102\penalty0 (3):\penalty0 527--540, 2001.

\bibitem[Petreanu et~al.(2009)Petreanu, Mao, Sternson, and
  Svoboda]{petreanu2009subcellular}
Leopoldo Petreanu, Tianyi Mao, Scott~M Sternson, and Karel Svoboda.
\newblock The subcellular organization of neocortical excitatory connections.
\newblock \emph{Nature}, 457\penalty0 (7233):\penalty0 1142--1145, 2009.

\bibitem[Magee(1998)]{magee1998dendritic}
Jeffrey~C Magee.
\newblock Dendritic hyperpolarization-activated currents modify the integrative
  properties of hippocampal {CA1} pyramidal neurons.
\newblock \emph{Journal of Neuroscience}, 18\penalty0 (19):\penalty0
  7613--7624, 1998.

\bibitem[Stuart and Spruston(1998)]{stuart1998determinants}
Greg Stuart and Nelson Spruston.
\newblock Determinants of voltage attenuation in neocortical pyramidal neuron
  dendrites.
\newblock \emph{Journal of Neuroscience}, 18\penalty0 (10):\penalty0
  3501--3510, 1998.

\bibitem[Harnett et~al.(2013)Harnett, Xu, Magee, and
  Williams]{harnett2013potassium}
Mark~T Harnett, Ning-Long Xu, Jeffrey~C Magee, and Stephen~R Williams.
\newblock Potassium channels control the interaction between active dendritic
  integration compartments in layer 5 cortical pyramidal neurons.
\newblock \emph{Neuron}, 79\penalty0 (3):\penalty0 516--529, 2013.

\bibitem[Harnett et~al.(2015)Harnett, Magee, and
  Williams]{harnett2015distribution}
Mark~T Harnett, Jeffrey~C Magee, and Stephen~R Williams.
\newblock Distribution and function of {HCN} channels in the apical dendritic
  tuft of neocortical pyramidal neurons.
\newblock \emph{Journal of Neuroscience}, 35\penalty0 (3):\penalty0 1024--1037,
  2015.

\bibitem[Turner et~al.(1991)Turner, Meyers, Richardson, and
  Barker]{turner1991site}
RW~Turner, DE~Meyers, TL~Richardson, and JL~Barker.
\newblock The site for initiation of action potential discharge over the
  somatodendritic axis of rat hippocampal {CA1} pyramidal neurons.
\newblock \emph{Journal of Neuroscience}, 11\penalty0 (7):\penalty0 2270--2280,
  1991.

\bibitem[Stuart and Sakmann(1994)]{stuart1994active}
Greg~J Stuart and Bert Sakmann.
\newblock Active propagation of somatic action potentials into neocortical
  pyramidal cell dendrites.
\newblock \emph{Nature}, 367\penalty0 (6458):\penalty0 69--72, 1994.

\bibitem[Schiller et~al.(1997)Schiller, Schiller, Stuart, and
  Sakmann]{schiller1997calcium}
Jackie Schiller, Yitzhak Schiller, Greg Stuart, and Bert Sakmann.
\newblock Calcium action potentials restricted to distal apical dendrites of
  rat neocortical pyramidal neurons.
\newblock \emph{The Journal of Physiology}, 505\penalty0 (3):\penalty0
  605--616, 1997.

\bibitem[Larkum et~al.(1999)Larkum, Zhu, and Sakmann]{larkum1999new}
Matthew~E Larkum, J~Julius Zhu, and Bert Sakmann.
\newblock A new cellular mechanism for coupling inputs arriving at different
  cortical layers.
\newblock \emph{Nature}, 398\penalty0 (6725):\penalty0 338--341, 1999.

\bibitem[Golding et~al.(2002)Golding, Staff, and
  Spruston]{golding2002dendritic}
Nace~L Golding, Nathan~P Staff, and Nelson Spruston.
\newblock Dendritic spikes as a mechanism for cooperative long-term
  potentiation.
\newblock \emph{Nature}, 418\penalty0 (6895):\penalty0 326--331, 2002.

\bibitem[Sj{\"o}str{\"o}m and H{\"a}usser(2006)]{sjostrom2006cooperative}
Per~Jesper Sj{\"o}str{\"o}m and Michael H{\"a}usser.
\newblock A cooperative switch determines the sign of synaptic plasticity in
  distal dendrites of neocortical pyramidal neurons.
\newblock \emph{Neuron}, 51\penalty0 (2):\penalty0 227--238, 2006.

\bibitem[Bittner et~al.(2015)Bittner, Grienberger, Vaidya, Milstein, Macklin,
  Suh, Tonegawa, and Magee]{bittner2015}
Katie~C Bittner, Christine Grienberger, Sachin~P Vaidya, Aaron~D Milstein,
  John~J Macklin, Junghyup Suh, Susumu Tonegawa, and Jeffrey~C Magee.
\newblock Conjunctive input processing drives feature selectivity in
  hippocampal {CA1} neurons.
\newblock \emph{Nature Neuroscience}, 18\penalty0 (8):\penalty0 1133, 2015.

\bibitem[Mel(1994)]{mel1994information}
Bartlett~W Mel.
\newblock Information processing in dendritic trees.
\newblock \emph{Neural computation}, 6\penalty0 (6):\penalty0 1031--1085, 1994.

\bibitem[K{\"o}rding and K{\"o}nig(2000)]{kording2000learning}
Konrad~P K{\"o}rding and Peter K{\"o}nig.
\newblock Learning with two sites of synaptic integration.
\newblock \emph{Network: Computation in Neural Systems}, 11\penalty0
  (1):\penalty0 25--39, 2000.

\bibitem[Poirazi et~al.(2003)Poirazi, Brannon, and Mel]{poirazi2003pyramidal}
Panayiota Poirazi, Terrence Brannon, and Bartlett~W Mel.
\newblock Pyramidal neuron as two-layer neural network.
\newblock \emph{Neuron}, 37\penalty0 (6):\penalty0 989--999, 2003.

\bibitem[Poirazi(2009)]{poirazi2009information}
Panayiota Poirazi.
\newblock Information processing in single cells and small networks: insights
  from compartmental models.
\newblock In \emph{AIP Conference Proceedings}, volume 1108, pages 158--167.
  American Institute of Physics, 2009.

\bibitem[Hendrickson et~al.(2011)Hendrickson, Edgerton, and
  Jaeger]{hendrickson2011capabilities}
Eric~B Hendrickson, Jeremy~R Edgerton, and Dieter Jaeger.
\newblock The capabilities and limitations of conductance-based compartmental
  neuron models with reduced branched or unbranched morphologies and active
  dendrites.
\newblock \emph{Journal of Computational Neuroscience}, 30:\penalty0 301--321,
  2011.

\bibitem[Urbanczik and Senn(2014)]{urbanczik2014learning}
Robert Urbanczik and Walter Senn.
\newblock Learning by the dendritic prediction of somatic spiking.
\newblock \emph{Neuron}, 81\penalty0 (3):\penalty0 521--528, 2014.

\bibitem[Guerguiev et~al.(2017)Guerguiev, Lillicrap, and
  Richards]{guerguiev2017towards}
Jordan Guerguiev, Timothy~P Lillicrap, and Blake~A Richards.
\newblock Towards deep learning with segregated dendrites.
\newblock \emph{Elife}, 6:\penalty0 e22901, 2017.

\bibitem[Sacramento et~al.(2018)Sacramento, Ponte~Costa, Bengio, and
  Senn]{sacramento2018dendritic}
Jo{\~a}o Sacramento, Rui Ponte~Costa, Yoshua Bengio, and Walter Senn.
\newblock Dendritic cortical microcircuits approximate the backpropagation
  algorithm.
\newblock \emph{Advances in Neural Information Processing Systems}, 31, 2018.

\bibitem[Haga and Fukai(2018)]{haga2018dendritic}
Tatsuya Haga and Tomoki Fukai.
\newblock Dendritic processing of spontaneous neuronal sequences for
  single-trial learning.
\newblock \emph{Scientific Reports}, 8\penalty0 (1):\penalty0 15166, 2018.

\bibitem[Richards and Lillicrap(2019)]{richards2019dendritic}
Blake~A Richards and Timothy~P Lillicrap.
\newblock Dendritic solutions to the credit assignment problem.
\newblock \emph{Current Opinion in Neurobiology}, 54:\penalty0 28--36, 2019.

\bibitem[Tzilivaki et~al.(2019)Tzilivaki, Kastellakis, and
  Poirazi]{tzilivaki2019challenging}
Alexandra Tzilivaki, George Kastellakis, and Panayiota Poirazi.
\newblock Challenging the point neuron dogma: {FS} basket cells as 2-stage
  nonlinear integrators.
\newblock \emph{Nature Communications}, 10\penalty0 (1):\penalty0 3664, 2019.

\bibitem[Jones and Kording(2021)]{jones2021might}
Ilenna~Simone Jones and Konrad~Paul Kording.
\newblock Might a single neuron solve interesting machine learning problems
  through successive computations on its dendritic tree?
\newblock \emph{Neural Computation}, 33\penalty0 (6):\penalty0 1554--1571,
  2021.

\bibitem[Milstein et~al.(2020)Milstein, Li, Bittner, Grienberger, Soltesz,
  Magee, and Romani]{milstein2020bidirectional}
Aaron~D Milstein, Yiding Li, Katie~C Bittner, Christine Grienberger, Ivan
  Soltesz, Jeffrey~C Magee, and Sandro Romani.
\newblock Bidirectional synaptic plasticity rapidly modifies hippocampal
  representations independent of correlated activity.
\newblock \emph{BioRxiv}, 2020.

\bibitem[Larkum(2022)]{larkum2022dendrites}
Matthew~E Larkum.
\newblock Are dendrites conceptually useful?
\newblock \emph{Neuroscience}, 489:\penalty0 4--14, 2022.

\bibitem[Mikulasch et~al.(2023)Mikulasch, Rudelt, Wibral, and
  Priesemann]{mikulasch2023error}
Fabian~A Mikulasch, Lucas Rudelt, Michael Wibral, and Viola Priesemann.
\newblock Where is the error? hierarchical predictive coding through dendritic
  error computation.
\newblock \emph{Trends in Neurosciences}, 46\penalty0 (1):\penalty0 45--59,
  2023.

\bibitem[Makarov et~al.(2023)Makarov, Pagkalos, and
  Poirazi]{makarov2023dendrites}
Roman Makarov, Michalis Pagkalos, and Panayiota Poirazi.
\newblock Dendrites and efficiency: Optimizing performance and resource
  utilization.
\newblock \emph{arXiv preprint arXiv:2306.07101}, 2023.

\bibitem[Hotelling(1936)]{hotelling1936relations}
Harold Hotelling.
\newblock Relations between two sets of variates.
\newblock \emph{Biometrika}, 28\penalty0 (3-4):\penalty0 321--377, 1936.

\bibitem[Arora et~al.(2017)Arora, Marinov, Mianjy, and
  Srebro]{arora2017stochastic}
Raman Arora, Teodor~Vanislavov Marinov, Poorya Mianjy, and Nati Srebro.
\newblock Stochastic approximation for canonical correlation analysis.
\newblock In \emph{Advances in Neural Information Processing Systems},
  volume~30, pages 4775--4784, 2017.

\bibitem[Bhatia et~al.(2018)Bhatia, Pacchiano, Flammarion, Bartlett, and
  Jordan]{bhatia2018gen}
Kush Bhatia, Aldo Pacchiano, Nicolas Flammarion, Peter~L Bartlett, and
  Michael~I Jordan.
\newblock Gen-{O}ja: Simple \& efficient algorithm for streaming generalized
  eigenvector computation.
\newblock In \emph{Advances in Neural Information Processing Systems},
  volume~31, pages 7016--7025, 2018.

\bibitem[Pehlevan et~al.(2020)Pehlevan, Zhao, Sengupta, and
  Chklovskii]{pehlevan2020neurons}
Cengiz Pehlevan, Xinyuan Zhao, Anirvan~M Sengupta, and Dmitri Chklovskii.
\newblock Neurons as canonical correlation analyzers.
\newblock \emph{Frontiers in Computational Neuroscience}, 14:\penalty0 55,
  2020.

\bibitem[Meng et~al.(2021)Meng, Chakraborty, and Singh]{meng2021online}
Zihang Meng, Rudrasis Chakraborty, and Vikas Singh.
\newblock An online {R}iemannian {PCA} for stochastic canonical correlation
  analysis.
\newblock \emph{Advances in Neural Information Processing Systems},
  34:\penalty0 14056--14068, 2021.

\bibitem[Gemp et~al.(2023)Gemp, Chen, and McWilliams]{gemp2023generalized}
Ian Gemp, Charlie Chen, and Brian McWilliams.
\newblock The generalized eigenvalue problem as a {N}ash equilibrium.
\newblock \emph{International Conference on Learning Representations}, 2023.

\bibitem[Boyd and Vandenberghe(2004)]{boyd2004convex}
Stephen Boyd and Lieven Vandenberghe.
\newblock \emph{Convex Optimization}.
\newblock Cambridge University Press, 2004.

\bibitem[Golkar et~al.(2020{\natexlab{a}})Golkar, Lipshutz, Bahroun, Sengupta,
  and Chklovskii]{golkar2020simple}
Siavash Golkar, David Lipshutz, Yanis Bahroun, Anirvan Sengupta, and Dmitri
  Chklovskii.
\newblock A simple normative network approximates local non-{H}ebbian learning
  in the cortex.
\newblock \emph{Advances in Neural Information Processing Systems},
  33:\penalty0 7283--7295, 2020{\natexlab{a}}.

\bibitem[Golkar et~al.(2020{\natexlab{b}})Golkar, Lipshutz, Bahroun, Sengupta,
  and Chklovskii]{golkar2020biologically}
Siavash Golkar, David Lipshutz, Yanis Bahroun, Anirvan~M Sengupta, and Dmitri~B
  Chklovskii.
\newblock A biologically plausible neural network for local supervision in
  cortical microcircuits.
\newblock \emph{arXiv preprint arXiv:2011.15031}, 2020{\natexlab{b}}.

\bibitem[F{\"o}ldi{\'a}k(1991)]{foldiak1991learning}
Peter F{\"o}ldi{\'a}k.
\newblock Learning invariance from transformation sequences.
\newblock \emph{Neural Computation}, 3\penalty0 (2):\penalty0 194--200, 1991.

\bibitem[Mitchison(1991)]{mitchison1991removing}
Graeme Mitchison.
\newblock Removing time variation with the anti-{H}ebbian differential synapse.
\newblock \emph{Neural Computation}, 3\penalty0 (3):\penalty0 312--320, 1991.

\bibitem[Wiskott and Sejnowski(2002)]{wiskott2002slow}
Laurenz Wiskott and Terrence~J Sejnowski.
\newblock Slow feature analysis: Unsupervised learning of invariances.
\newblock \emph{Neural Computation}, 14\penalty0 (4):\penalty0 715--770, 2002.

\bibitem[Berkes and Wiskott(2005)]{berkes2005slow}
Pietro Berkes and Laurenz Wiskott.
\newblock Slow feature analysis yields a rich repertoire of complex cell
  properties.
\newblock \emph{Journal of vision}, 5\penalty0 (6):\penalty0 9--9, 2005.

\bibitem[Franzius et~al.(2007)Franzius, Sprekeler, and
  Wiskott]{franzius2007slowness}
Mathias Franzius, Henning Sprekeler, and Laurenz Wiskott.
\newblock Slowness and sparseness lead to place, head-direction, and
  spatial-view cells.
\newblock \emph{PLoS computational biology}, 3\penalty0 (8):\penalty0 e166,
  2007.

\bibitem[Barlow(1989)]{barlow1989unsupervised}
Horace~B Barlow.
\newblock Unsupervised learning.
\newblock \emph{Neural Computation}, 1\penalty0 (3):\penalty0 295--311, 1989.

\bibitem[Atick(1992)]{atick1992could}
Joseph~J Atick.
\newblock Could information theory provide an ecological theory of sensory
  processing?
\newblock \emph{Network: Computation in Neural Systems}, 3\penalty0
  (2):\penalty0 213--251, 1992.

\bibitem[Bell and Sejnowski(1996)]{bell1996edges}
Anthony Bell and Terrence~J Sejnowski.
\newblock Edges are the ``independent components'' of natural scenes.
\newblock \emph{Advances in Neural Information Processing Systems}, 9:\penalty0
  831--837, 1996.

\bibitem[Bell and Sejnowski(1997)]{bell1997independent}
Anthony~J Bell and Terrence~J Sejnowski.
\newblock The ``independent components'' of natural scenes are edge filters.
\newblock \emph{Vision Research}, 37\penalty0 (23):\penalty0 3327--3338, 1997.

\bibitem[Lewicki(2002)]{lewicki2002efficient}
Michael~S Lewicki.
\newblock Efficient coding of natural sounds.
\newblock \emph{Nature Neuroscience}, 5\penalty0 (4):\penalty0 356--363, 2002.

\bibitem[Cardoso(1989)]{cardoso1989source}
J-F Cardoso.
\newblock Source separation using higher order moments.
\newblock In \emph{International Conference on Acoustics, Speech, and Signal
  Processing}, pages 2109--2112. IEEE, 1989.

\bibitem[Abid et~al.(2018)Abid, Zhang, Bagaria, and Zou]{abid2018exploring}
Abubakar Abid, Martin~J Zhang, Vivek~K Bagaria, and James Zou.
\newblock Exploring patterns enriched in a dataset with contrastive principal
  component analysis.
\newblock \emph{Nature Communications}, 9\penalty0 (1):\penalty0 1--7, 2018.

\bibitem[McLachlan(2005)]{mclachlan2005discriminant}
Geoffrey~J McLachlan.
\newblock \emph{Discriminant analysis and statistical pattern recognition}.
\newblock John Wiley \& Sons, 2005.

\bibitem[de~Cheveign{\'e} and Parra(2014)]{de2014joint}
Alain de~Cheveign{\'e} and Lucas~C Parra.
\newblock Joint decorrelation, a versatile tool for multichannel data analysis.
\newblock \emph{Neuroimage}, 98:\penalty0 487--505, 2014.

\bibitem[Pehlevan and Chklovskii(2014)]{pehlevan2014hebbian}
Cengiz Pehlevan and Dmitri~B Chklovskii.
\newblock A {H}ebbian/anti-{H}ebbian network derived from online non-negative
  matrix factorization can cluster and discover sparse features.
\newblock In \emph{Signals, Systems and Computers, 2014 48th Asilomar
  Conference on}, page 769{\textendash}775. IEEE, 2014.

\bibitem[Bahroun and Soltoggio(2017)]{bahroun2017online}
Yanis Bahroun and Andrea Soltoggio.
\newblock Online representation learning with single and multi-layer {H}ebbian
  networks for image classification.
\newblock In \emph{International Conference on Artificial Neural Networks},
  pages 354--363. Springer, 2017.

\bibitem[Sengupta et~al.(2018)Sengupta, Tepper, Pehlevan, Genkin, and
  Chklovskii]{sengupta2018manifold}
Anirvan Sengupta, Mariano Tepper, Cengiz Pehlevan, Alexander Genkin, and Dmitri
  Chklovskii.
\newblock Manifold-tiling localized receptive fields are optimal in
  similarity-preserving neural networks.
\newblock In \emph{Advances in Neural Information Processing Systems}, 2018.

\bibitem[Qin et~al.(2021)Qin, Mudur, and Pehlevan]{qin2021contrastive}
Shanshan Qin, Nayantara Mudur, and Cengiz Pehlevan.
\newblock Contrastive similarity matching for supervised learning.
\newblock \emph{Neural Computation}, 33\penalty0 (5):\penalty0 1300--1328,
  2021.

\bibitem[Pehlevan et~al.(2017{\natexlab{b}})Pehlevan, Mohan, and
  Chklovskii]{pehlevan2017blind}
Cengiz Pehlevan, Sreyas Mohan, and Dmitri~B Chklovskii.
\newblock Blind nonnegative source separation using biological neural networks.
\newblock \emph{Neural Computation}, 29\penalty0 (11):\penalty0 2925--2954,
  2017{\natexlab{b}}.

\bibitem[Golkar et~al.(2022)Golkar, Te{\c{s}}ileanu, Bahroun, Sengupta, and
  Chklovskii]{golkar2022constrained}
Siavash Golkar, Tiberiu Te{\c{s}}ileanu, Yanis Bahroun, Anrivan Sengupta, and
  Dmitri~B Chklovskii.
\newblock Constrained predictive coding as a biologically plausible model of
  the cortical hierarchy.
\newblock \emph{Advances in Neural Information Processing Systems}, 35, 2022.

\bibitem[Friedrich et~al.(2021)Friedrich, Golkar, Farashahi, Genkin, Sengupta,
  and Chklovskii]{friedrich2021neural}
Johannes Friedrich, Siavash Golkar, Shiva Farashahi, Alexander Genkin, Anirvan
  Sengupta, and Dmitri~B Chklovskii.
\newblock Neural optimal feedback control with local learning rules.
\newblock \emph{Advances in Neural Information Processing Systems},
  34:\penalty0 16358--16370, 2021.

\bibitem[Lipshutz et~al.(2023{\natexlab{b}})Lipshutz, Kashalikar, Farashahi,
  and Chklovskii]{lipshutz2022linear}
David Lipshutz, Aneesh Kashalikar, Shiva Farashahi, and Dmitri~B Chklovskii.
\newblock A linear discriminant analysis model of imbalanced associative
  learning in the mushroom body compartment.
\newblock \emph{PLoS Computational Biology}, 19\penalty0 (2):\penalty0
  e1010864, 2023{\natexlab{b}}.

\bibitem[Duong et~al.(2023)Duong, Lipshutz, Heeger, Chklovskii, and
  Simoncelli]{duong2023adaptive}
Lyndon~R Duong, David Lipshutz, David Heeger, Dmitri~B Chklovskii, and Eero~P
  Simoncelli.
\newblock Adaptive whitening in neural populations with gain-modulating
  interneurons.
\newblock \emph{International Conference on Machine Learning}, 2023.

\end{thebibliography}

\appendix

\end{document}